\documentclass[a4paper,11pt]{article}
\pdfoutput=1 

\usepackage{jinstpub} 
\usepackage{graphicx}
\usepackage{textcomp}
\usepackage[english]{babel}
\usepackage{hyperref} 
\usepackage{xcolor}
\usepackage{pagecolor}
\usepackage{siunitx}
\usepackage{lineno}
\usepackage{csquotes}
\usepackage[backend=biber, style=ieee,language=auto]{biblatex}
\DeclareSourcemap{ 
    \maps{
        \map{ 
            \step[fieldsource=url,
                  match=\regexp{\{\\\_\}|\{\_\}|\\\_},
                  replace=\regexp{\_}]
        }
        \map{ 
            \step[fieldsource=url,
                  match=\regexp{\{\$\\sim\$\}|\{\~\}|\$\\sim\$},
                  replace=\regexp{\~}]
        }
        \map{ 
            \step[fieldsource=url,
                  match=\regexp{\{\\\x{26}\}},
                  replace=\regexp{\x{26}}]
        }
    }
}

\title{\boldmath A modular and flexible data acquisition system for a cosmic rays detector network }

\author[a,1]{G. T. Saito,\note{Corresponding author.}}
\author[a]{M. A. L. Leite,}
\author[a]{R. Menegasso,}
\author[a]{M. K. Kuriyama,}
\author[a]{M. G. Munhoz}
\author[a]{and R. E. de Paula}

\affiliation[a]{Instituto de Física da Universidade de São Paulo,\\São Paulo, Brazil}

\emailAdd{g.saito@cern.ch}

\abstract{In this paper, we describe a modular data acquisition system developed as the foundation  of a cosmic ray detector network. Each detector setup (henceforth referred as a station) is composed of an independent hardware device that can be controlled and read-out through the Internet. This device is designed to acquire and process the signal of up to eight different detector planes. Each of these  detector planes uses plastic scintillator slabs that are optically coupled to silicon photomultipliers (SiPM). Within a single station, different geometries and plane orientations are possible using the same baseline design. The main readout is based on a programmable system-on-a-chip (PSoC), a flexible and re-configurable commodity hardware that is used to implement the trigger and timing logic. A Time to Digital Converter (TDC) is used to determine the precise timing of the event relative to a GPS timing signal and to estimate the signal amplitude through the Time-over-Threshold (ToT) method. An auxiliary set of sensors provide environmental information and station detector planes orientation that, together with  other operation data, are periodically sent to a server using the MQTT protocol.
Data is cached using an in-memory database for online monitoring and further persisted into a SQL database for offline analysis. The server framework is  based in software application containers allowing easy replication of the server infrastructure.}


\proceeding{TWEPP 2021 Topical Workshop on Electronics for Particle Physics\\
  20 to 24 September 2021\\
  Online}

\begin{document}

\maketitle
\flushbottom

\section{Introduction}
The detection of cosmic rays using simple apparatuses for quantitative data-taking have been explored over the years by several initiatives around the world for outreach and experimental High Energy Physics instrumentation \textcolor{black}{teaching \cite{Ruchti2002Quarknet:U.s.a.} \cite{cosmicpi} \cite{fokkema2012hisparc}.} The possibility to connect geographically dispersed stations synchronized by a GPS timing \textcolor{black}{signal \cite{Berns:2003zfa}} allows for a larger detection area suitable for the identification of high energy cosmic rays showers. In order for such system to be deployed in a high school or science museum environment (hence for outreach purposes), it must be safe (no flammable gases or high voltages) and ideally low cost, so the network can comprise as many stations as possible. The use of plastic scintillators, SiPMs and commodity electronic hardware allows the project to fulfill this objective. The readout, timing and trigger implementation supports a variety of geometries and even other cosmic rays detection methods \textcolor{black}{(e.g. Cherenkov)} that can be easily deployed in the system by a simple reconfiguration of the hardware. This enables a variety of experiments to be performed with a single design, thus reducing the complexity and \textcolor{black}{cost} of the system construction and operation.

\section{The Cosmic Ray Station}

\begin{figure}[htbp]
\centerline{\includegraphics[width=0.7\linewidth,trim=0 0 0 15cm, clip]{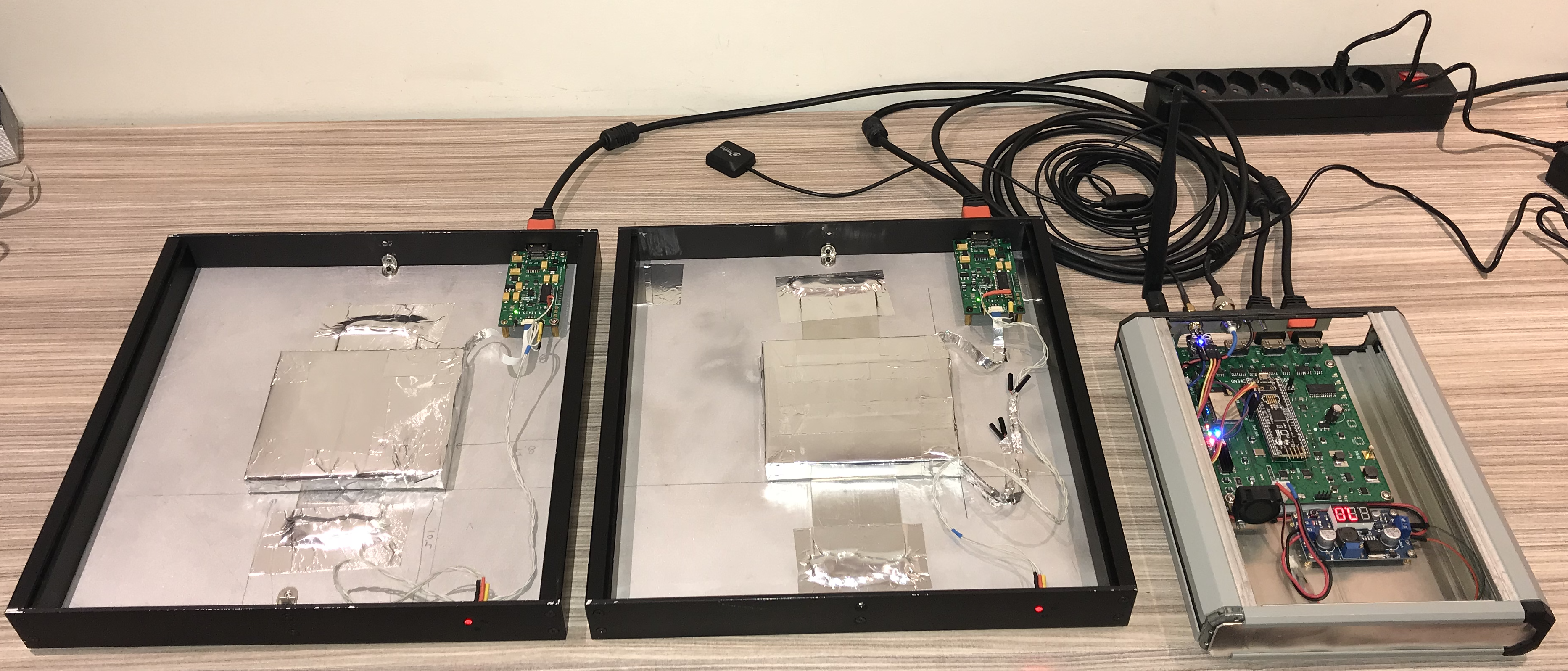}}
\caption{Photo showing 2 Frontend \textcolor{black}{modules (left) and a Backend module (right)}.}
\label{fig:photo}
\end{figure}

Each station is an autonomous hardware (and firmware) unity  that detects cosmic ray events and transmits the raw data to the software stack through the Internet. These stations are composed of up to four
{\em Frontend} modules, in which the scintillators, SiPMs and the analog section of the electronics are located in separate boards and a {\em Backend} module aggregating the trigger, timing, data acquisition, event building and network communication functions (Fig. \ref{fig:photo}).

\subsection{Frontend detection board}

The current system uses $\SI{3.9}{\milli\metre}\times\SI{3.9}{\milli\metre}$ SiPMs sensors \textcolor{black}{(Broadcom AFBR-S4N44C013 \cite{broadcom44})} that can be installed as an array of up to four sensors mounted on  dedicated PCBs. The SiPMs are attached to a plastic scintillator slab \textcolor{black}{(BC408)} using optical grease. The scintillators in use range from $\SI{150}{\milli\metre}\times\SI{150}{\milli\metre}\times\SI{10}{\milli\metre}$ to $\SI{400}{\milli\metre}\times\SI{400}{\milli\metre}\times\SI{10}{\milli\metre}$ . The current design is able  to use \textcolor{black}{scintillators} slabs as thick as \SI{20}{\milli\meter}. An EEPROM device  installed in each detection board (read-out by I2C protocol) allows for matching the inventory of built boards with the boards deployed in a given site.   The system is sealed in a light-proof aluminum container and the connection to the readout electronics is made using a strip line flat cable. Two LEDs can be pulsed in order to provide a light signal for debug and calibration.

\subsection{Frontend readout board}

Each Frontend  holds a two-channel  voltage amplifier followed by a discriminator circuit. Each channel can sum-up the signal from two SiPMs or be ganged together to sum the signal from four sensors when a thicker scintillator slab is used. A minimum detection system can comprise only one Frontend module with two separate channels, and detect the passage of cosmic rays  by requiring the coincidence of the two signals.  

The Frontend readout board also holds a microcontroller with digital-to-analog (DAC) converters which are used to set the threshold voltage of the two discriminators independently and to fine adjust the SiPM bias voltage. The discriminated signal is sent to a Backend module, where all timing and acquisition are performed.  This adjustment is made using a high voltage amplifier in a non-inverting setup, so the different SiPMs sensors can be biased by a low noise adjustable power supply. 

Each Frontend board deploys an inertial measurement unit \textcolor{black}{(IMU ICM-20948 \cite{ICM-20948TDK})} sensor with a 3-axis accelerometer, gyroscope, and compass, so the orientation of the planes relative to the earth surface can always be determined. Alongside the IMU sensor,  an environmental sensor \textcolor{black}{(Bosch BME680 \cite{GasSensortec})} is used to measure temperature, humidity, and atmosphere pressure. \textcolor{black}{The} data can then be further  used to correlate the cosmic rays flux measurements to the environmental and operating conditions.

\subsection{Backend module}

\begin{figure}
    \vspace{1cm}
    \centering
    \includegraphics[width=0.8\linewidth]{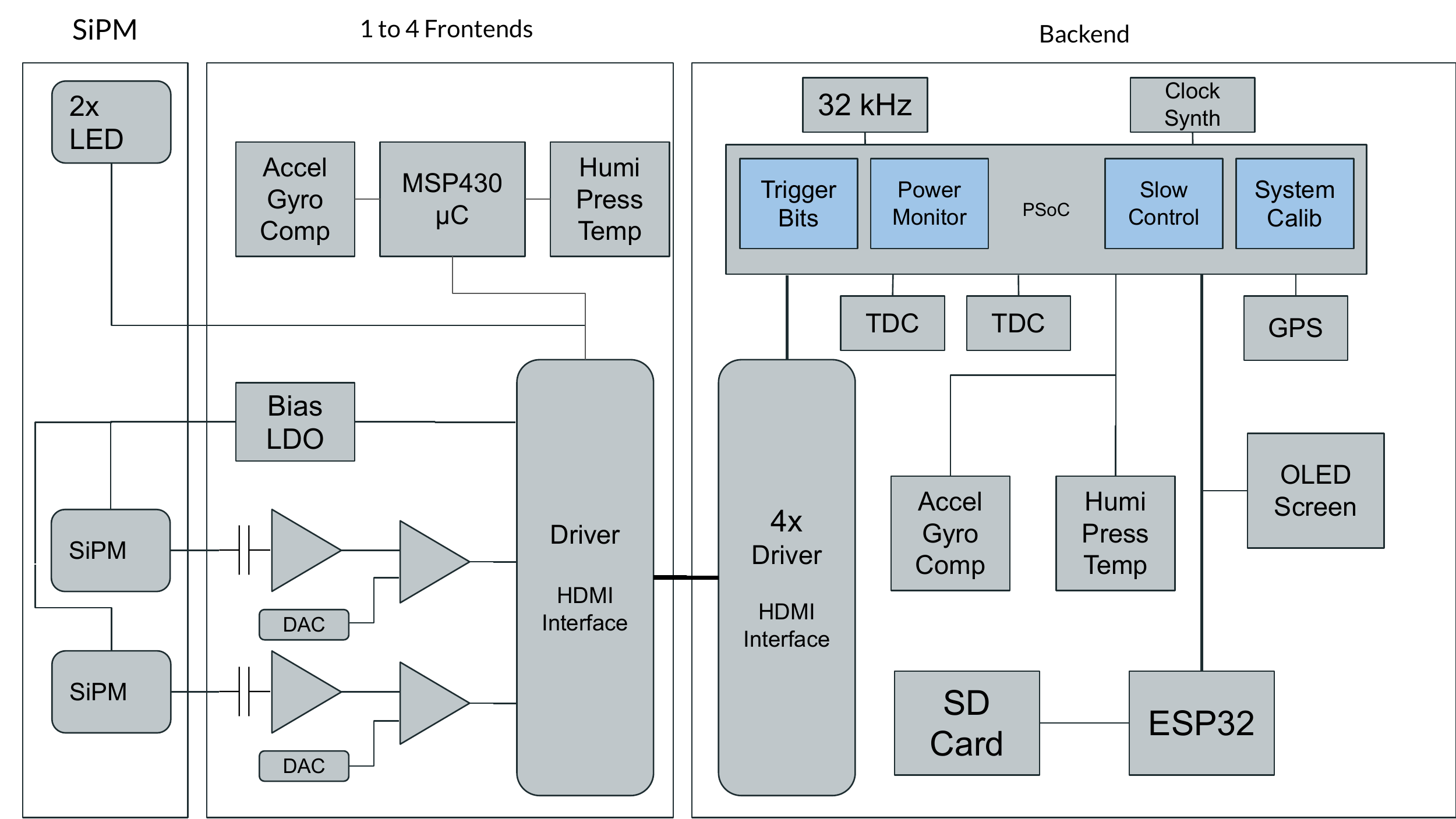}
    \caption{\\\textit{Block Diagram of a detector showing the PCBs and their internal components.}}
    \label{fig:blockdiagram}
\end{figure}

The Backend module is a single board responsible for  the trigger, timing and data acquisition. The power for all the station and network connectivity is also provided though the Backend. Up to four Frontend modules can be connected to a single Backend module through a commodity  HDMI cable in which the four differential lanes are used to carry two digitized SiPM and two LED calibration signals. \textcolor{black}{Fig. \ref{fig:blockdiagram} describes these connections and the peripherals around the PCBs.}

\begin{figure}[htbp]
\centerline{\includegraphics[width=.4\linewidth]{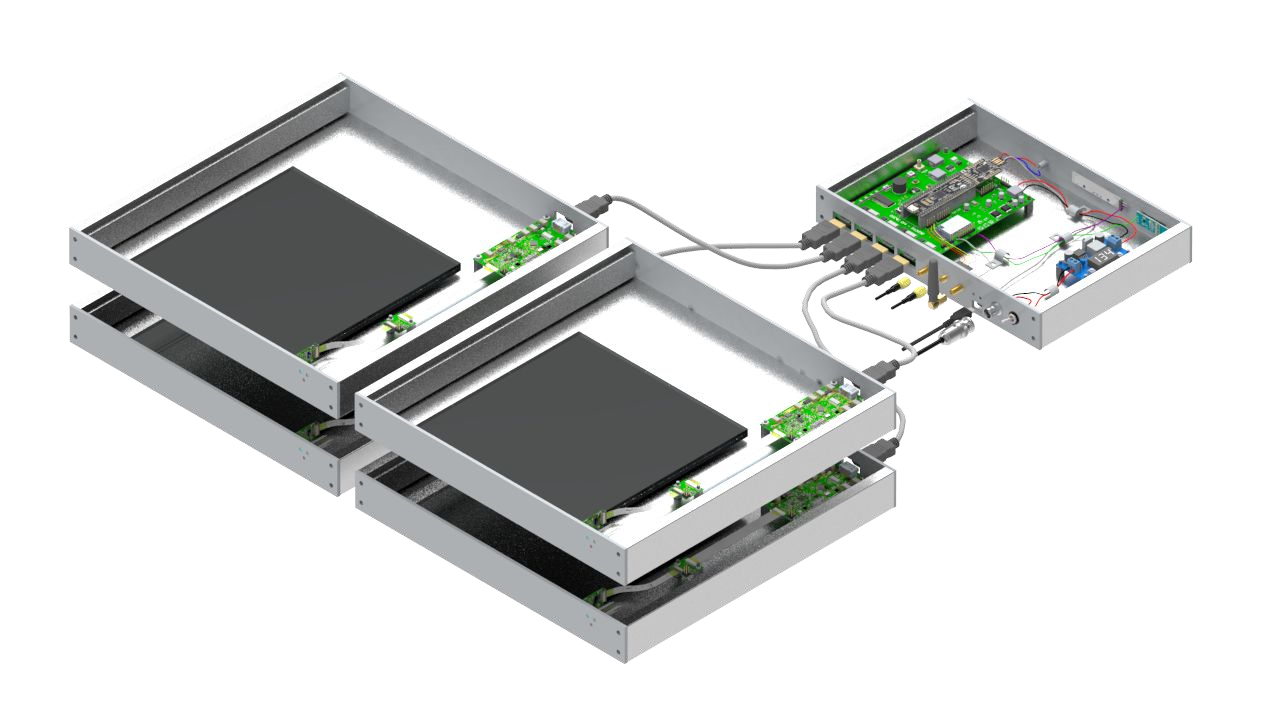}
\includegraphics[width=.4\linewidth]{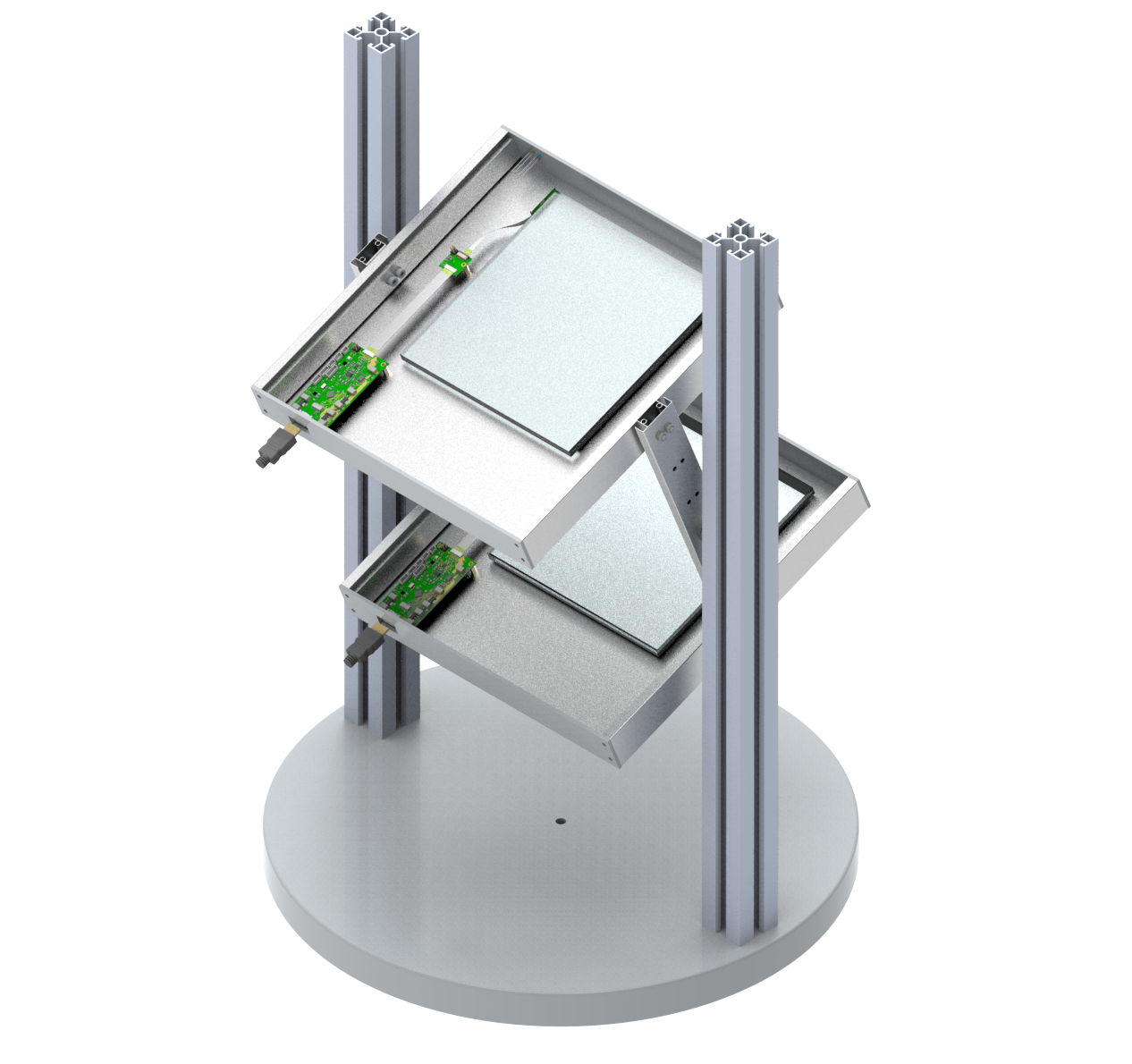}}
\caption{Some of the possible arrangements for the cosmic ray detector planes. The trigger configuration can be selected before the operation to reflect the geometrical placement of the planes.}
\label{fig:render}
\end{figure}

A Cypress Semiconductor Programmable-System-on-a-Chip (Cypress PSoC5LP \cite{psoc}) performs most of the data processing of the station. Using the PSoC programmable digital and analog \textcolor{black}{circuitry, the signals from the Frontend modules are routed to a trigger predefined look-up table, and the trigger signal is generated from its output. Up to 5 different combinatory operations between the inputs from the detector are possible, providing a reconfigurable trigger decision logic without the complexity of an external FPGA. Therefore, the user can choose to configure the trigger with any logical operation between any channels based on the spacial disposition of detection planes in the station, as an example, Fig. \ref{fig:render} shows a doubled detection area and telescope-like configuration of detection planes.}

Two external Time to Digital Converter (TI TDC7200) are used for \textcolor{black}{precise (better than 50ns \cite{Berns:2003zfa})} timing determination between the trigger signal and the GPS timing synchronization signal. This allows for the off-line synchronization of events acquired by independent stations within the GPS timing resolution. The current GPS module \textcolor{black}{(uBlox Neo-6 \cite{neo6}) offers} a final resolution of 30ns, but the very high timing resolution of the TDC (in the range of 50ps) enables the future use of GPS modules deploying advanced timing techniques, and hence providing a better timing resolution. For some  applications, the SiPM discriminated signals can be internally \textcolor{black}{routed in the PSoC to the TDC input channels for measuring the width of the pulse and through the use of a Time-over-Threshold (ToT) method, estimate the pulse amplitude and energy.}

Finally, all the information collected by the PSoC5LP can be either stored in a local flash memory or sent to a WiFi SoC module \textcolor{black}{(Espressif Systems ESP32 \cite{ESP32Systems})} that, connected to a local Wi-Fi network, will relay the information to the Internet. The station uses the \textcolor{black}{MQTT protocol \cite{MQTTMessaging}} to receive configuration data and to send the acquired data to a server. In the server, data is cached in an {\em in-memory} database \textcolor{black}{(Redis \cite{Redis})} for on-line monitoring of all connected stations and persisted in a SQL \textcolor{black}{(Oracle MySQL \cite{MySQL})} database for offline analysis.

\subsection{Initialization, calibration and operation}

During the initialization, the operating conditions of the station can be either read from the non-volatile memory in the PSoC5L or be set remotely by the network. This allows for fine tuning the detector bias and discriminator thresholds over time without physical intervention in the station. A calibration procedure was devised to establish the offsets in the time measurements and to verify the system integrity, and it is always performed before any data-taking. A series of test trigger patterns can also be injected  by either pulsing the LEDs or by digitally forcing signals in the PSoC5LP internal logic in order to debug the station operation. 

The normal operation is usually performed in an unattended fashion after choosing  a specific  trigger decision configuration. The data taking is sliced in time or in number of events in {\em Cosmic blocks}, \textcolor{black}{so the analysis can filter out whole Cosmic blocks that where taken when the detector operating conditions are out of specifications. At the beginning of each Cosmic block, all the environmental, voltages and currents of the system and the software state are read out by the Backend module and sent to the database.} A data quality procedure can later on check for non-conforming Cosmic blocks, tagging then as not good for analysis. Using the timestamp information of the event, an offline analysis can search for timing coincidences among different nearby stations, \textcolor{black}{indicating detection of a large cosmic rays showers.}   

\section{Data Management and Analysis}

Intended as an outreach focused project, the user interface allows many educational activities, ranging from just observing the detection rate from a fixed station to the combination of several stations or angular muon distribution.
The infrastructure is built as a \textcolor{black}{Docker stack \cite{EmpoweringDocker}}, with each service running in its own container. Each station has a dedicated topic in our \textcolor{black}{MQTT Broker \cite{MQTTMessaging}} and a custom Python script offloads all messages transmitted in these topics to a \textcolor{black}{Redis cache \cite{Redis}, a fast in-memory database.} This script is also responsible for some data reconstruction and the \textcolor{black}{Database (SQL) storage \cite{MySQL}}. Two parallel services are available for the user interface: the first is a \textcolor{black}{Grafana panel \cite{Grafana:Labs}}, intended as a read-only web page showing all monitor information predefined by the team; the second one is a \textcolor{black}{Jupyter Notebook \cite{ProjectHome} service}, a web-based interactive notebook capable of running Python code with read access to the Database through the network.

\section{Results}
\begin{figure}[ht!]
    \vspace{1cm}
    \centering
    \includegraphics[width=0.8\linewidth]{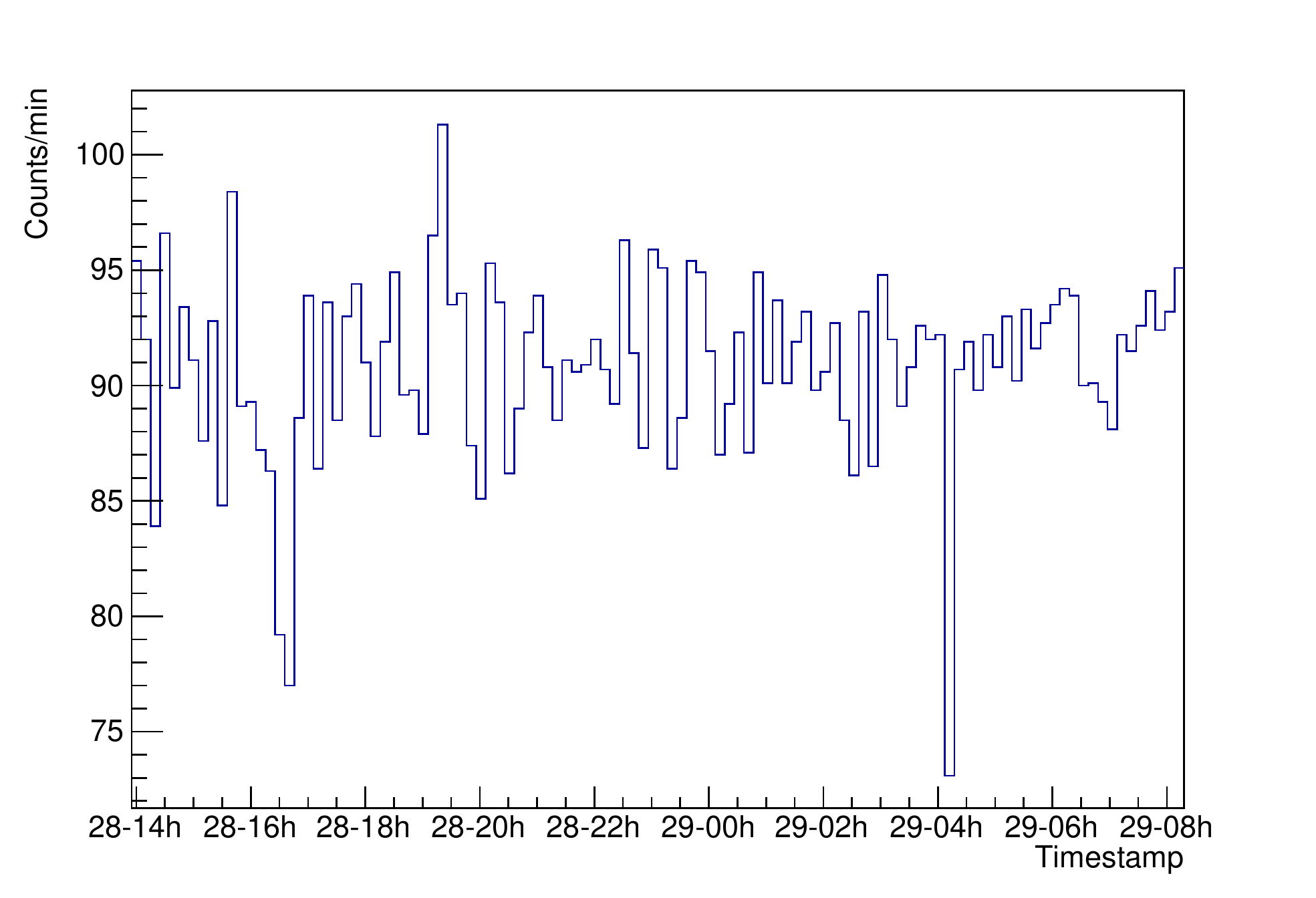}
    \caption{Plot showing the number of counts per minute of data taking over 2 days with an AND trigger logic between 4 stacked detection planes of $\SI{150}{\milli\metre}\times\SI{150}{\milli\metre}$ using BC408 \SI{4}{\milli\metre} slabs.}
    \label{fig:hist}
\end{figure}

\textcolor{black}{
Data taking with the detector started with a configuration of 2 Frontend modules made of two $\SI{150}{\milli\metre}\times\SI{150}{\milli\metre}\times\SI{4}{\milli\metre}$ slabs of plastic scintillator (BC408) each. Fig. \ref{fig:hist} shows a preliminary result taken in the city of São Paulo, Brazil, at an altitude of approximately \SI{760}{\meter}. The detector was located inside a building under a \SI{40}{\centi\meter} thick concrete ceiling. Using an AND coincidence as a trigger between all 4 detection planes, the station registered an average of 90 events per minute.
}


\acknowledgments

We acknowledge support from RENAFAE and CNPq (Proc. 440519/2019-5) for this project.



\printbibliography

@misc{psoc,
    title = {{32-bit Arm{\textregistered} Cortex{\textregistered}-M3 PSoC{\textregistered} 5LP}},
    year = {2020},
    number = {PSoC{\textregistered} 5LP: CY8C52LP},
    url = {https://www.cypress.com/products/32-bit-arm-cortex-m3-psoc-5lp}
}

@misc{cosmicpi,
    title = {{Cosmic Pi | The cosmic ray detector on your desktop}},
    url = {http://cosmicpi.org/}
}

@article{Berns:2003zfa,
    title = {{GPS Time Synchronization in School-Network Cosmic Ray Detectors}},
    year = {2004},
    journal = {IEEE Trans. Nucl. Sci.},
    author = {Berns, H -G. and Burnett, T H and Gran, R and Wilkes, R J},
    editor = {Valentine, J D},
    pages = {848--853},
    volume = {51},
    doi = {10.1109/TNS.2004.829368},
    arxivId = {physics/0311079}
}

@misc{neo6,
    title = {{NEO-6 u-blox 6 GPS Modules}},
    year = {2011},
    number = {NEO-6},
    url = {https://www.u-blox.com/sites/default/files/products/documents/NEO-6_DataSheet_(GPS.G6-HW-09005).pdf}
}

@misc{broadcom44,
    title = {{NUV-HD Single Silicon Photo Multiplier}},
    year = {2019},
    number = {AFBR-S4N44C013},
    url = {https://docs.broadcom.com/doc/AFBR-S4N44C013-DS}
}

@article{fokkema2012hisparc,
    title = {{The Hisparc cosmic ray experiment: data acquisition and reconstruction of shower direction}},
    year = {2012},
    author = {Fokkema, DBRA},
    doi = {10.3990/1.9789036534383}
}

@misc{EmpoweringDocker,
    title = {{Empowering App Development for Developers | Docker}},
    url = {https://www.docker.com/}
}

@misc{ESP32Systems,
    title = {{ESP32 Wi-Fi {\&} Bluetooth MCU I Espressif Systems}},
    url = {https://www.espressif.com/en/products/socs/esp32}
}

@misc{GasSensortec,
    title = {{Gas Sensor BME680 | Bosch Sensortec}},
    url = {https://www.bosch-sensortec.com/products/environmental-sensors/gas-sensors/bme680/}
}

@misc{Grafana:Labs,
    title = {{Grafana: The open observability platform | Grafana Labs}},
    url = {https://grafana.com/}
}

@misc{ICM-20948TDK,
    title = {{ICM-20948 | TDK}},
    url = {https://invensense.tdk.com/products/motion-tracking/9-axis/icm-20948/}
}

@misc{MQTTMessaging,
    title = {{MQTT - The Standard for IoT Messaging}},
    url = {https://mqtt.org/}
}

@misc{MySQL,
    title = {{MySQL}},
    url = {https://www.mysql.com/}
}

@misc{ProjectHome,
    title = {{Project Jupyter | Home}},
    url = {https://jupyter.org/}
}

@inproceedings{Ruchti2002Quarknet:U.s.a.,
    title = {{Quarknet: a Particle Physics Program of Education and Outreach in the U.s.a.}},
    year = {2002},
    booktitle = {7th International Conference on Advanced Technology and Particle Physics},
    author = {Ruchti, R C},
    doi = {10.1142/9789812776464_0079}
}

@misc{Redis,
    title = {{Redis}},
    url = {https://redis.io/}
}







\end{document}